\documentclass[american,prl,showpacs,reprint]{revtex4-1}
\usepackage[utf8]{inputenc}
\usepackage{mathtools}
\usepackage{amsmath}
\usepackage{amssymb}
\usepackage{mathdots}
\usepackage{graphicx}

\makeatletter

\DeclareTextSymbolDefault{\textquotedbl}{T1}

\usepackage{gensymb}

\makeatother

\usepackage{babel}
\begin{document}
\title{Adsorbate induced manipulation of 1D atomic nanowires:\\
Soliton mediated degradation of long-range order in the Si(553)-Au
system}
\affiliation{Department of Physics and Center for Nanointegration (CENIDE), University
of Duisburg-Essen, 47057 Duisburg, Germany}
\author{B.~Hafke}
\email{bernd.hafke@uni-due.de}

\author{T.~Witte}
\author{M.~Horn-von~Hoegen}
\affiliation{Department of Physics and Center for Nanointegration (CENIDE), University
of Duisburg-Essen, 47057 Duisburg, Germany}
\date{\today}
\begin{abstract}
Deposition of Au on vicinal Si(553) surfaces results in the self-assembly
of one-dimensional (1D) Au atomic wires. Charge transfer from the
Au wire to the Si step edge leads to a chain of Si dangling-bond orbitals
with a long range ordered threefold periodicity along the steps and
finite interchain interaction perpendicular to the steps. Employing
spot-profile analysis low-energy electron diffraction (SPA-LEED) we
observed a broadening of spot width with time, reflecting the degradation
of Si dangling bond chain and Au wire length. Adsorbates were identified
as primary source for spot broadening. They force the generation of
solitons and anti-solitons which immediately destroy the long-range
order along and perpendicular to the steps, respectively. From the
temporal evolution of broadening, we conclude that the Au wires are
less reactive to adsorption than the Si dangling bond chains.
\end{abstract}
\maketitle

\subsection*{Introduction and motivation}

The assembly of single atoms to wires has recently attracted much
attention, because such one-dimensional (1D) structures exhibit unique
physical properties \citep{yeom1999instability,himpsel2001one}. The
intrinsic low dimensionality of such systems causes pronounced instabilities
in the electronic and structural degrees of freedom resulting in cooperative
phenomena like the formation of charge density waves or spin density
waves, depending on the interaction strength between electron system
and lattice \citep{snijders2010colloquium,Crain2004}. Additionally,
the interactions within the surface layer is crucial in the long-range
ordering of the wires. Highly ordered arrays of atomic wires can be
fabricated by self-assembly through the deposition of metal atoms
on well defined vicinal semiconductor surfaces \citep{Segovia,Aulbach2016}.
Varying the substrate's vicinality allows the manipulation of interwire
coupling and results in the modification of physical properties like
the surface conductance \citep{okino2007transport,Edler2017}. The
understanding of such fundamental interaction mechanisms may lead
to the utilization of 1D metallic wires as a toolbox for the fabrication
of 2D and 1D systems with individually tailored properties \citep{Aulbach2016}.
However, for these applications the fragility of such systems is crucial,
but has not been studied to far and thus warrants further investigation.

In this context, Au nanowires on the Si(553) surface have attracted
a lot of attention. The formation of Au double-strands with dimerization
along the steps is observed \citep{PhysRevB.82.075426,PhysRevB.81.115436}.
Adjacent wires do not show a long-range correlation across the steps,
but only minor electronic interactions with the Si atoms \citep{PhysRevB.96.081406,Braun2018}.
The step edge is formed by a Si honeycomb-like structure. Due to charge
redistribution within the surface layer every third Si step-edge atom
along the steps exhibits an unsaturated dangling-bond (DB) orbital.
It has been reported that this DB has a single electron occupation
which leads to the intrinsic full spin polarization of this DB state
\citep{erwin2010intrinsic,Aulbach2016}. The condensation of these
spins into an (anti-)ferromagnetic DB array could be ruled out through
the observation of ordering of DBs perpendicular to the steps: A centered-like
geometry of the Si DB unit cell results in a magnetically frustrated
array that favors a 2D quantum spin liquid phase rather than (anti-)ferromagnetic
ordering \citep{hafke2016two}.

The latest modification of the structural model predicts every third
Si DB orbital to be unoccupied, which leads to the complete extinction
of spin-interaction between the adjacent Si DB chains \citep{Braun2018}.
The much stronger interstep interaction between the Si DB chains compared
to the Au wires can be attributed to Coulomb interactions of the partially
charged DBs.

Here we report, that ordering of the DB chains and Au wires in 1D
and 2D is fragile and readily affected by adsorbates. In a microscopic
picture, solitons induce translational phase shifts in the Si DB chain
and Au wire structure, respectively, resulting in gradual loss of
long range order.  Employing spot-profile analysis low-energy electron
diffraction (SPA-LEED) \citep{scheithauer1986new,horn1999growth},
we quantitatively describe the reduction of Si chain and Au wire lengths
along the steps and the loss of long-range order perpendicular to
the steps.

\subsection*{Methods and sample system}

The experiments were performed under ultra-high vacuum conditions
(UHV) at a base pressure below $1\cdot10^{-10}\text{ mbar}$. The
substrate was cut from a \textit{n}-type Si(553) wafer (phosphorus
doped, $0.01\mathrm{\ \Omega cm}$) with a nominal miscut of $<0.5\,{^\circ}$
and mounted on a liquid helium cryostat. Prior to deposition the sample
was degassed at $650\,^{\circ}\text{C}$ for several hours. The sample
was cleaned in several short flash-anneal cycles by heating via direct
current to $1250\,^{\circ}\text{C}$. An amount of $0.48\text{ ML}$
(monolayer, referred to the atomic density of a Si(111) surface, i.e.,
$1\text{ ML}=7.83\cdot10^{14}\text{ atoms}/\text{cm}^{2}$) Au was
deposited from an electron beam heated graphite crucible \citep{kury2005compact}
at a substrate temperature of $650\,^{\circ}\text{C}$. The deposition
was followed by a post annealing step \citep{PhysRevLett.111.137203}
at $850\,^{\circ}\text{C}$ for several seconds and subsequent rapid
cooling to a temperature of $T=60\text{ K}$ or $T=80\text{ K}$,
respectively.

\onecolumngrid

\begin{figure}
	\begin{centering}
	\includegraphics[width=0.9\textwidth]{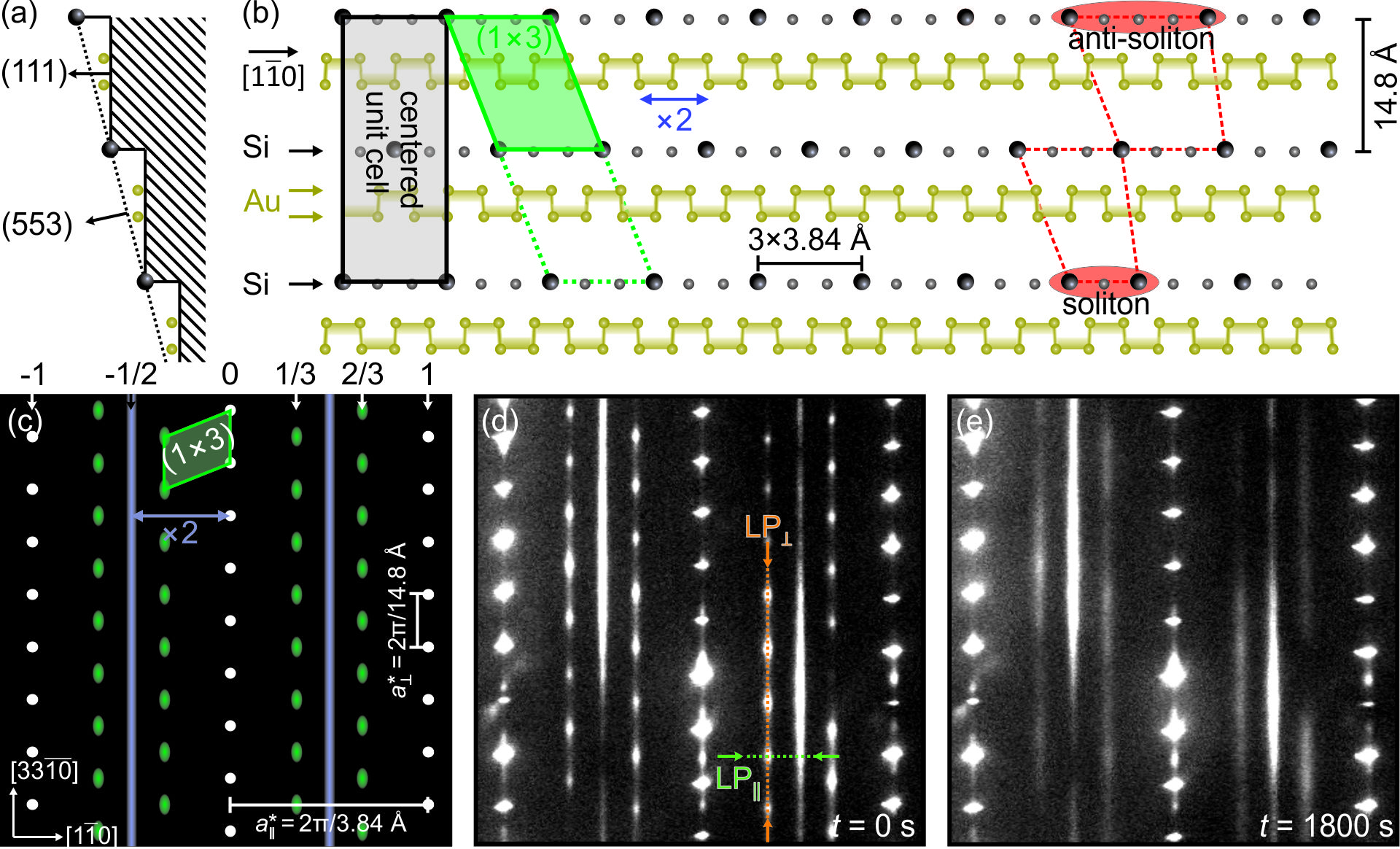}
	\par\end{centering}
	\caption{\label{fig:Diffpattern}(a,b) Schematic structure model of the Si(553)-Au
		surface. Side view in (a) reveals terraces with (111) orientation.
		Top view of (b) shows the dimerized Au atoms as golden spheres, the
		saturated Si DBs as small gray spheres, and the unsaturated Si DBs
		as large gray spheres. The geometry of the Si DB array can either
		be described by the primitive ($1\times3$)-unit cell in green or
		by the centered unit cell in gray. In the right part a soliton and
		an anti-soliton is indicated as red ellipse, respectively. (c) Schematic
		LEED pattern of the perfect structure. The rows of integer order spots
		are indicated as white circles, the $\times3$ spots as green ellipses,
		and the $\times2$ streaks in blue lines.(d) LEED pattern at $t=0\text{ s}$
		directly after preparation shows perfect agreement with the ideal
		pattern from (c). (e) LEED pattern at $t=1800$ s after preparation
		shows clear broadening of the $\times3$ spots. All patterns were
		recorded at $E=150$ eV and a temperature of $T=78$ K.}
\end{figure}
\twocolumngrid

Figure~\ref{fig:Diffpattern}~(a,b) show a schematic structure model
of the Si(553)-Au system. Each step has a height of $3.14\text{\,Å}$
and is separates by terraces of $14.4\text{\,Å}$ width with a $(111)$
orientation, as indicated in the side view of Fig.~\ref{fig:Diffpattern}~(a).
Accordingly, the Si(553) surface exhibits a miscut of about $12.3\,{^\circ}$
with respect to the $(111)$ orientation. The steps in the $[33\overline{10}]$
direction are separated by $a_{\perp}=14.8\text{\,Å}$, as sketched
in the top view of Fig.~\ref{fig:Diffpattern}~(b). The atomic distance
along the step direction of the commensurate structure is $a_{0}=3.84\text{\,Å}$.
The Au double-strand wires (golden spheres) decorate the middle of
the steps along the $[1\overline{1}0]$ direction. The dimerization
of the Au wires is indicated by alternating long and short interatomic
distances, respectively. The Si step edge atoms are sketched as gray
spheres. Every third Si atom along the steps (larger gray spheres)
is embedded into a 2D array of unsaturated DBs forming a centered-like
unit cell (gray shaded rectangle).

A schematic diffraction pattern resulting from this surface is shown
in Fig.~\ref{fig:Diffpattern}~(c). Three dense rows of integer
order spots (white dots and indicated by -1, 0 and 1) reflect diffraction
from the periodic step train of the Si(553) substrate. The narrow
distance of $2\pi/14.8\text{\,Å}$ between the integer order spots
along the $[33\overline{10}]$ direction is caused by the step separation
$a_{\perp}$. The wide separation of $2\pi/3.84\text{\,Å}$ reflects
the spacing $a_{0}$ of the Si atoms along the step. Two additional
periodicities along the $[1\overline{1}0]$ direction, i.e., between
the rows of integer order spots can be identified. First, the streaks
(light blue line) located half between the rows of integer order spots
are caused by the twofold ($\times2$) periodicity of the dimerized
Au wires. The dimerization between Au wires of adjacent terraces is
almost uncorrelated and thus a streak instead of spots appears. Second,
rows of weaker spots with a separation of $2\pi/14.8\text{\,Å}$
(elongated green dots) arise at positions $1/3$ and $2/3$ between
the rows of integer order spots. These superstructure spots reflect
the threefold ($\times3$) periodicity of the DB chain along the Si
step edge. The primitive $(1\times3)$-unit cell and the centered
unit cell are depicted in green and gray, respectively.

\begin{figure}
	\begin{centering}
		\includegraphics[width=0.95\columnwidth]{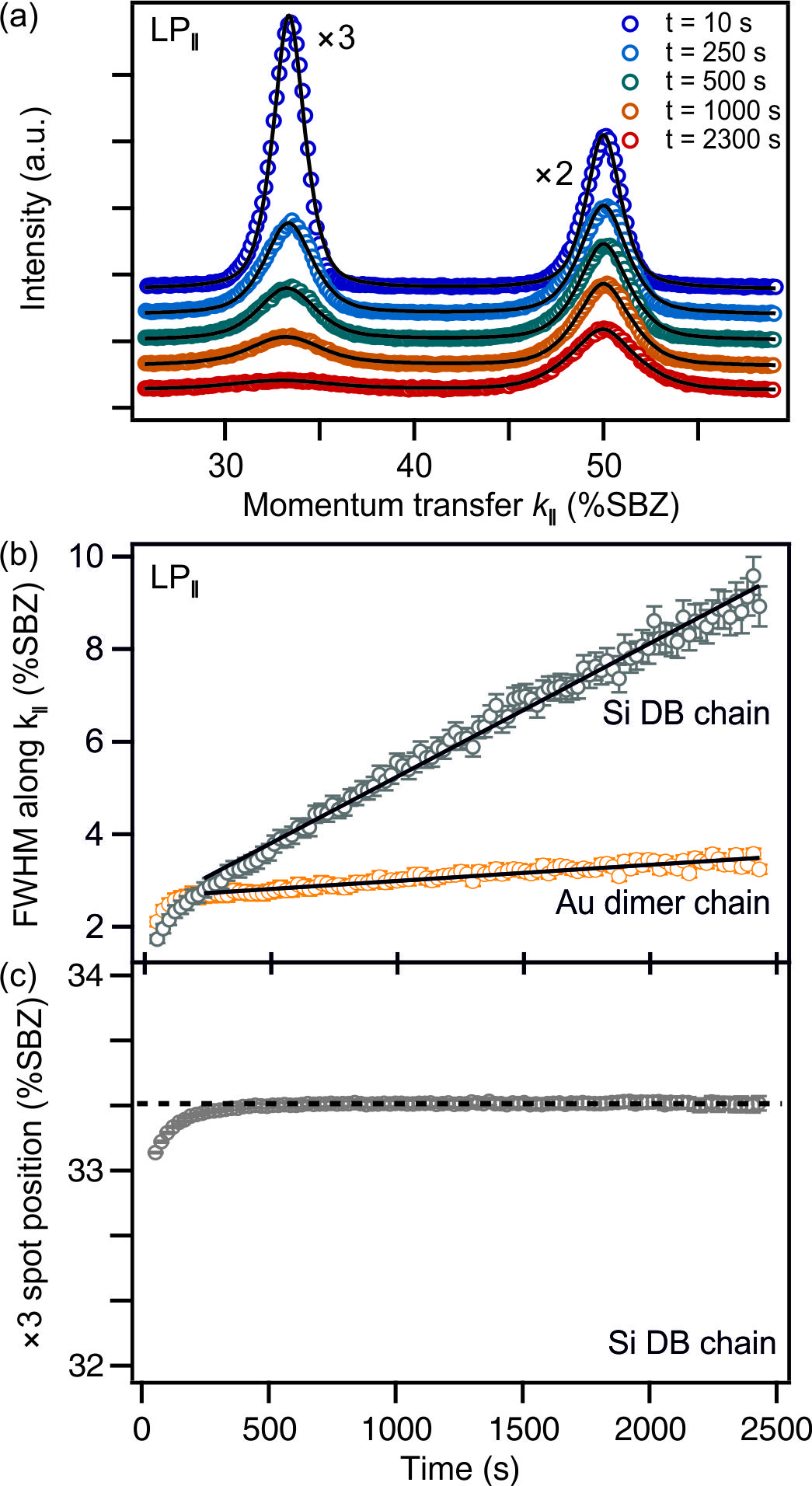}
		\par\end{centering}
	\caption{\label{fig:LPs}(a) Time series of spot profiles recorded at $T=60$
		K perpendicular to $\times3$ spots and $\times2$ streaks (LP indicated
		as green dotted line in Fig.~\ref{fig:Diffpattern}~(d)) indicating
		the evolution of long-range order along the Si DB chains and Au wires
		with time, respectively. LPs are shifted vertically for clarity. (b)
		The FWHM ($100\text{ \%SBZ}=2\pi/3.84\,\text{Å}$) of $\times3$
		spots (from the Si DB chains) and $\times2$ streaks (from the Au
		wires) as determined from Lorentzian fits to the LPs shown in (a)
		in dependence of adsorption time. Black solid lines indicate linear
		FWHM increase with time. (c) The Position of the $\times3$ spot is
		independent on adsorption time for $t>200$~s (black dashed constant
		line).}
\end{figure}

\subsection*{Results}

After preparation of the Si(553)-Au surface and rapid cool down to
$80\text{ K}$ we observe a continuous degradation of the initial
LEED pattern with time. Figure~\ref{fig:Diffpattern}~(d, e) depict
the LEED patterns as prepared ($t=0\text{ s}$) and after $t=1800\text{ s}$
of adsorption, respectively. The patterns are shown in a linear intensity
scale to emphasize the relevant features of the superstructure spots
and thus the integer order spots are overexposed. While the LEED pattern
in Fig.~\ref{fig:Diffpattern}~(d) at $t=0\text{ s}$ exhibits clear
spots at $1/3$ position, these spots are strongly broadened in Fig.~\ref{fig:Diffpattern}~(e)
at $t=1800\text{ s}$ both in the $[1\overline{1}0]$ and in the $[33\overline{10}]$
direction.

Short flash-anneal cycles of few seconds duration up to $850\,^{\circ}\text{C}$
result in desorption of adsorbates, no depletion or loss of Au coverage
occurs \citep{horn1999induced}, and the surface structure can be
refreshed. The LEED pattern exhibits sharp spots again, i.e., the
initial surface preparation is completely restored. As we do not observe
any degradation of the LEED pattern through electron irradiation,
we have to conclude evidence for adsorbate (from residual gas) induced
broadening of the superstructure $\times3$ spots and $\times2$ streaks.

In order to quantify the increase of spot broadening we recorded -
as function of adsorption time -- intensity line profiles perpendicular
and parallel to the streaks in the diffraction pattern, i.e., the
$[1\overline{1}0]$ direction (indicated through $\text{LP}_{\parallel}$
in Fig.~\ref{fig:Diffpattern}~(d) by the dotted green line), and
perpendicular to the DB chains, i.e., the $[33\overline{10}]$ direction
(indicated by $\text{\text{LP}}_{\perp}$ in Fig.~\ref{fig:Diffpattern}~(d)
by the dotted orange line), respectively. All LPs are recorded at
a base temperature of $T=60\text{ K}$. Fig.~\ref{fig:LPs}~(a)
depicts selected line profiles along $\text{LP}_{\parallel}$ as function
of adsorption time. These line profiles through the $\times3$ spot
and $\times2$ streak reflect the order along the Si DB chains and
along the dimerized double-strand Au wires, respectively.

The profiles for the $\times3$ spot and $\times2$ streak are both
described by Lorentzian functions
\begin{equation}
\mathcal{L}(k,\kappa)\propto\frac{1}{\kappa^{2}+(k-k_{0})^{2}}
\end{equation}
 with peak position $k_{0}$ and full width at half maxima (FWHM)
of the surface Brillouin-zone (SBZ), $\text{FWHM}=2\kappa$, without
any traces of a sharp central spike (see the solid lines in Fig.~\ref{fig:LPs}~(a)).
The evolution of the related full width at half maxima (FWHM) for
the $\times3$ spot and $\times2$ streak in dependence of adsorption
time is shown in Fig.~\ref{fig:LPs}~(b). The curves exhibit a linear
increase of FWHM with time of $(0.173\pm0.002)\text{ \%SBZ/min}$
and $(0.021\pm0.001)\text{ \%SBZ/min}$ for the $\times3$ spot and
$\times2$ streak, respectively. For the very initial regime of adsorption
the slope is even steeper and may be attributed to a crossover from
2D order to 1D order \citep{Hafke_PRL}. The position of the $\times3$
spot is plotted in Fig.~\ref{fig:LPs}~(c). Except for a small deviation
at the initial regime of adsorption the position stays constant at
1/3 of the distance between the rows of integer order spots.

Fig.~\ref{fig:Corrlength}~(a) depicts selected spot profiles $\text{LP}_{\perp}$
along the row of $\times3$ spots in $[33\overline{10}]$ direction
(reflecting the real space order between adjacent Si DB chains) as
function of adsorption time. These profiles are described by a sum
of equidistant Lorentzian functions with identical FWHM and without
any traces of central spikes. With increasing adsorption time the
spots rapidly broaden until at $t=1000$ s only a streak at 1/3 position
between the row of integer order spots remains (see diffraction pattern
in Fig.~\ref{fig:Diffpattern}~(e)). The evolution of the corresponding
FWHM in $[33\overline{10}]$ direction for the row of $\times3$ spots
in dependence of adsorption time is shown in Fig.~\ref{fig:Corrlength}~(b).
The curve exhibit an initial increase of FWHM with time of $(2.70\pm0.08)\text{ \%SBZ/min}$.
Thus, the spot broadening of the $\text{LP}_{\perp}$ direction increases
faster with exposure time to residual gas than for the $\text{LP}_{\parallel}$
direction.

\subsection*{Discussion}

In diffraction spot profiles with Lorentzian shape are indicative
for the presence of anti-phase translational domains which exhibit
a geometric size distribution \citep{henzler1978quantitative,lent1984diffraction,Pukite1985,Wollschlaeger2007}.
The lateral and/or vertical phase shift may thereby arise from steps
associated with islands on flat surfaces \citep{lent1984diffraction,Pukite1985},
steps at vicinal surfaces (\citep{Foelsch1997,Tegenkamp2002,Wollschlaeger2007},
or domain boundaries of superstructure domains \citep{Nagao1998,hild2000induced}.

First, the mechanism for adsorbate induced spot broadening on the
basis of the $\times3$ spots arising from the Si DB chains at the
step edges is discussed. As we do not observe a systematic variation
of the spot profiles with the vertical momentum transfer, we can exclude
morphological changes of the surface during adsorption. Instead, we
interpret the Lorentzian spot profile as a signature of ordered Si
DB chains with $\times3$ periodicity with finite lengths (1D domains).
Then, the length distribution of such 1D domains of ordered DB chains
is a geometric distribution that is indicative for a Markov process
\citep{spadacini1983markovian}. Our experiment is sensitive to a
maximum length limited by the instrumental transfer width of $\xi\approx30\text{ nm}$.
The chains are confined by adsorbates originating from the residual
gas. These adsorbates stick at the highly reactive Si step edges \citep{Edler2017}
and pinpoint the positions of the $\times3$ ordering of the Si DB
chain with respect to the neighboring chains. Inevitably, this leads
to the generation of heavy or light zero dimensional anti-phase domain
boundaries, i.e., of solitons or anti-solitons, respectively \citep{Hafke_PRL}.
Both types of solitons are schematically visualized in the right part
of Fig.~\ref{fig:Diffpattern}~(b) highlighted through red ellipses.
The solitons and anti-solitons result in a shift of the $\times3$
Si DB by $\pm a_{0}$ along the steps. As the position of the $\times3$
spots does not change during adsorption we conclude, that both, solitons
and anti-solitons are generated with the same probability independent
on the density of adsorbates.

\begin{figure}
	\begin{centering}
		\includegraphics[width=0.9\columnwidth]{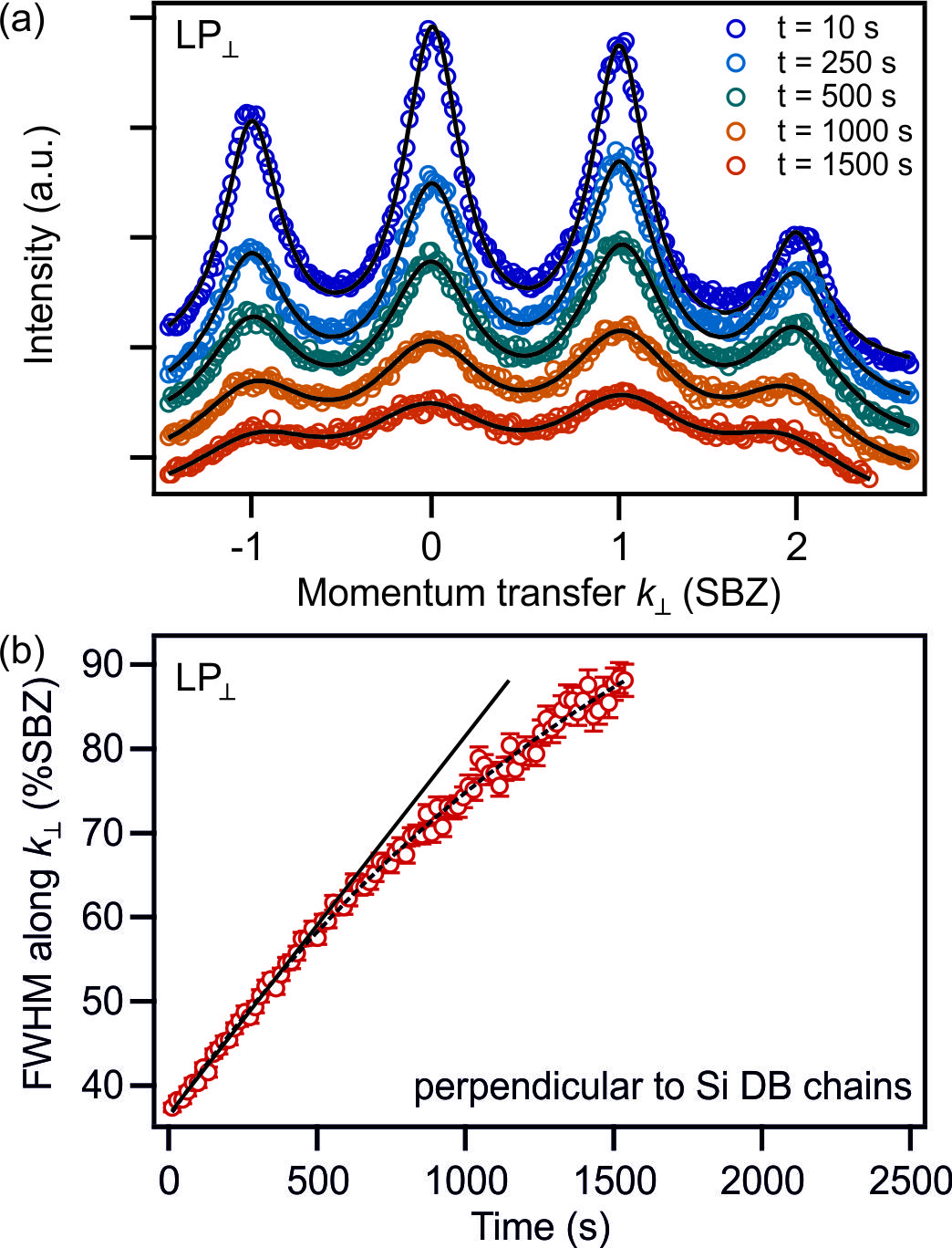}
		\par\end{centering}
	\caption{\label{fig:Corrlength}(a) Series of line profiles recorded at $T=60$
		K along the row of $\times3$ spots (LP indicated as orange dotted
		line in Fig.~\ref{fig:Diffpattern}~(d)) indicating the evolution
		of long-range order of Si DB chains perpendicular to the steps with
		time. LPs are shifted vertically for clarity. (b) The FWHM ($100\text{ \%SBZ}=2\pi/14.8\,\text{Å}$)
		as determined from a series of equidistant Lorentzian functions fitted
		to the line profiles (indicated as black curves) shown in (a). The
		solid black line reflects the linear increase of FWHM in the beginning
		of adsorption. The dashed curve provides a guide to the eye.}
\end{figure}

The absence of a sharp central spike in the spot profiles indicates
the loss of the long-range $\times3$ order in the Si DB chains already
with the adsorption of the very first adsorbates. Thus, the entire
Si DB chain is shifted along the steps and is only confined by other
pinning defects or solitons. Solitons are mobile at finite temperatures
\citep{erwin2013silicon,Hafke_PRL}. As the motion of a DB by $\pm a_{0}$
is initiated by electron hopping without the necessity of bond breaking
only a small energy barrier has to be overcome. This motion is accompanied
by a relaxation of the related Si atomic positions \citep{erwin2010intrinsic,Braun2018}.
Thus, we expect this process to be much faster than the time between
two independent adsorption events.

Vertical to the chains, the FWHM increases faster than along the chains.
Thus, the long-range centered order between the DB chains of neighboring
steps is destroyed faster. The adsorbate-induced generation of solitons
or anti-solitons in each DB chain leads not only to phase shifts within
the DB chain, but also to extended phase shifts between adjacent DB
chains along the chain direction. The shift of the DBs by one single
adsorbate already destroys the centered correlation to both neighboring
chains and is therefore particularly destructive. With increasing
adsorbate density and thus arbitrary sequence of $\times3$ ordered
DB chains, the phase correlation between neighboring chains is lifted
extremely efficiently. Finally, the transition to streaks in diffraction
(Fig.~\ref{fig:Corrlength}~(a) line profile at $t=1500$ s) reflects
the associated complete loss of correlation between adjacent DB chains.

The observations for the $\times2$ streak arising from the dimerization
of the Au wire is interpreted in a similar way as we have done for
the $\times3$ Si DB chains. Adsorbates from residual gas on the Au
wires trigger a flip of dimerization along the terraces. This phase
shift of dimerization of $\Delta\phi=\pi$ is the fundamental excitation
of these wires and requires to overcome an energy barrier. Compared
with the Si DB chains, where both solitons and anti-solitons are equally
present and the number of possible states is three, the Au wires exhibit
only two states and just one type of phase boundary, i.e., solitons
only. In contrast to the Si DB chain the associated phase shift requires
bond breaking and reformation of Au-Au bonds between the dimerized
Au atoms within the wire \citep{Braun2018}. The local perturbation
through an adsorbate is strong enough to provide the necessary energy.
As in the case for the Si DB chains the $\times2$ streaks arising
from the Au wires do not exhibit a sharp central spike. Instead, the
continuous increase of FWHM of the Lorentzian shaped $\times2$ streak
also indicates the immediate loss of long-range order of dimerized
Au atoms.

Since the rise of FWHM (golden data points in Fig.~\ref{fig:LPs}~(b))
for the Au wires is less steep than for the DB chains, we conclude
that the Au wire is less reactive to adsorption. This is also supported
by DFT calculations \citep{Edler2017} where the preferred adsorption
sites for $\text{O}_{2}$ are located at the Si step edge and not
on the Au wires located in the middle of the terrace. From the vastly
different slopes of the FWHM attributed to the Si DB chains and Au
wires, we conclude that adsorbates on the Si DB chains have no influence
to the long-range order of the Au wires, because otherwise the slope
of change in FWHM would be the same for both adsorption sites.

\subsection*{Summary and Conclusions}

We studied adsorbate-induced degradation of long-range 1D and 2D ordered
atomic wires in the Si(553)-Au system by means of SPA-LEED. A strong
increase of spot width as function of adsorption time is indicative
for a reduction of the lengths of $\times3$ ordered Si DB chains
and $\times2$ dimerized Au wires length along the steps, respectively.
This is explained by adsorbates forcing the generation of solitons
or anti-solitons. From the absence of central spikes in the spot profiles
we conclude that the long-range order of both Si DB chains and Au
wires, respectively, is gradually degraded starting with the first
adsorbates. The spot positions of the $\times3$ spots do not change
with increasing adsorbate density proving that solitons and anti-solitons
are generated with equal probability. We found a much lower reactivity
of the Au wires compared to the Si DB chains. This corroborates the
findings of preferred adsorption sites to the Si step edge deduced
from transport measurements and DFT \citep{Edler2017}. The mechanisms
of soliton/anti-soliton generation and soliton motion in the case
of adsorption is closely related to thermal activation of soliton/anti-soliton
pairs in that system \citep{Hafke_PRL}. Microscopically, the energies
for these fundamental excitations which are mediated by the adsorbates
have to be in the same order as the energies necessary for thermal
excitations. We expect these processes which lead to the loss of long-range
order also occur in other atomic wire systems of the Ge/Si($hhk$)-Au
family.

\subsection*{Acknowledgments}
\begin{acknowledgments}
Financial support via funding of the Deutsche Forschungsgemeinschaft
(DFG, German Research Foundation) -- Projektnummer 278162697 --
SFB1242 project C03 \textquotedbl Driven phase transitions at surfaces:
initial dynamics, hidden states and relaxation\textquotedbl{} is gratefully
acknowledged. Financial support from the DFG FOR1700 research unit
``metallic nanowires on the atomic scale: Electronic and vibrational
coupling in real world systems'' is gratefully acknowledged.
\end{acknowledgments}


\end{document}